\documentclass[10pt,twocolumn]{article}

\usepackage{graphicx}
\usepackage{algorithm,algorithmic}
\usepackage{amsmath}
\usepackage{multirow}
\usepackage{float}
\usepackage{url}
\usepackage[none]{hyphenat}
\usepackage{authblk}

\usepackage{afterpage}

\newcommand\TTT{\rule{0pt}{2.2ex}}

\begin{document}
\newcounter{cntr1}
\newcounter{cntr2}
\emergencystretch 3em

\title{Multi-Objective Optimisation of Damper Placement for
Improved Seismic Response in Dynamically Similar Adjacent Buildings}

\author[1]{Mahesh B. Patil}
\author[2]{U. Ramakrishna}
\author[2]{S. C. Mohan}
\affil[1]{Department of Electrical Engineering, Indian Institute of Technology Bombay}
\affil[2]{Department of Civil Engineering, BITS Pilani Hyderabad Campus}

\maketitle

\begin{abstract}
Multi-objective optimisation of damper placement in dynamically
symmetric adjacent buildings is considered with identical viscoelastic
dampers used for vibration control. First, exhaustive search is
used to describe the solution space in terms of various quantities
of interest such as maximum top floor displacement, maximum floor
acceleration, base shear, and interstorey drift. With the help of examples,
it is pointed out that the Pareto fronts in these problems contain
a very small number of solutions. The effectiveness of two commonly
used multi-objective evolutionary algorithms, viz., NSGA-II and
MOPSO, is evaluated for a specific example.
\end{abstract}

\section{Introduction}
The use of energy dissipation devices to reduce structural
vibrations arising due to earthquake excitations has been
reported extensively (see \cite{domenico2019} for a review).
The effect of damper placement as well as damper parameters
has been studied in detail. The techniques used for the above
purpose can be broadly divided into two categories:
(a)\,parametric study
\cite{bhaskararao2006}-%
\nocite{patel2010}%
\cite{bharti2014}
in which damper parameters or positions are varied in a systematic
manner and their effect on the response of interest obtained,
(b)\,optimisation of damper parameters and/or positions with respect
to one or more objective functions
\cite{singh2002}-%
\nocite{wongprasert2004}%
\nocite{park2004}%
\nocite{matsagar2005}%
\nocite{liu2005}%
\nocite{makita2007}%
\nocite{lagaros2007}%
\nocite{lavan2009}%
\nocite{bigdeli2012}%
\nocite{aydin2013}%
\nocite{hadi2015}%
\nocite{park2015}%
\nocite{mastali2016}%
\nocite{aydin2017}%
\nocite{barraza2017}%
\nocite{bogdanovic2019}%
\nocite{cetin2019}%
\nocite{puthanpurayil2019}%
\cite{rahmani2019}.

\begin{figure*}[!ht]
\centering
\scalebox{0.9}{\includegraphics{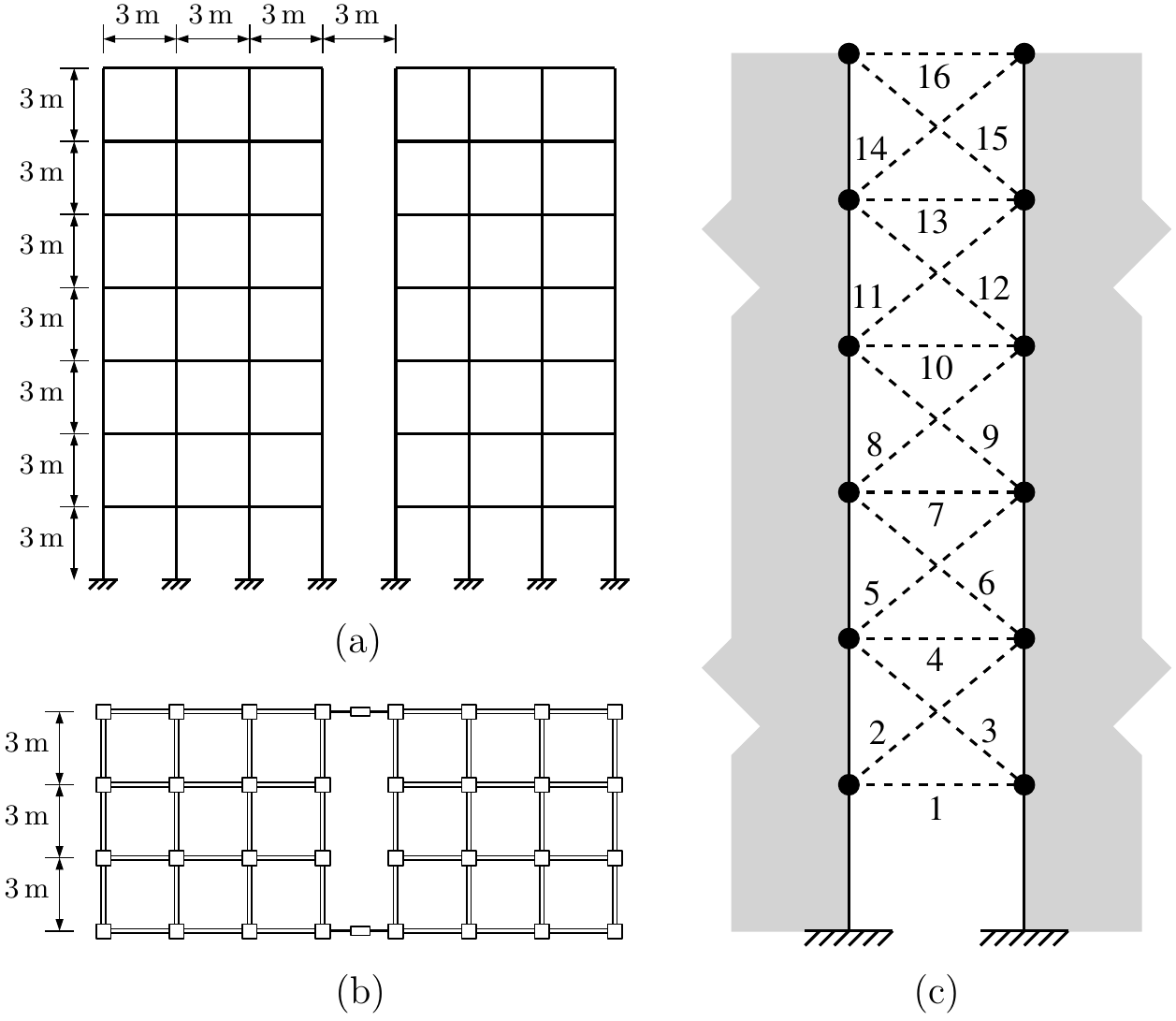}}
\vspace*{-0.2cm}
\caption{Two dynamically similar adjacent buildings:
(a)\,Elevation, (b)\,plan, (c)\,schematic representation.}
\label{fig_bldg}
\end{figure*}

In this paper, we consider a specific case, viz., reduction of seismic
response for two dynamically similar adjacent buildings (DSABs)
on a rigid foundation using viscoelastic dampers\,%
\cite{bhaskararao2006},%
\cite{patel2010}, as shown in Fig.~\ref{fig_bldg}.
Each dashed line in
Fig.~\ref{fig_bldg}\,(c)
corresponds to two dampers (at either end, as shown in
Fig.~\ref{fig_bldg}\,(b)). The equation of motion can be
written in the following form\,\cite{patel2010}:
\begin{equation}
{\bf{M}}\,{\ddot{\bf{X}}}
+ ({\bf{C}} + {\bf{C}}_D)\,{\dot{\bf{X}}}
+ {\bf{K}}\,{\bf{X}}
= - {\bf{M}}\,{\bf{I}}\,{\ddot{x_g}}.
\label{eq_motion}
\end{equation}

A detailed description of Eq.~\ref{eq_motion} is given in
\cite{patel2010}. The vector {\bf{X}} contains the relative
displacement of each floor (in each of the two buildings)
due to the ground acceleration excitation given by
${\ddot{x_g}}$. Throughout this paper, we have considered
the 1940 El Centro ground acceleration.

Eq.~\ref{eq_motion} is solved using Newmark's
method\,\cite{rajasekaran} to yield
${\bf{X}}(t)$,
${\dot{\bf{X}}}(t)$, an
${\ddot{\bf{X}}}(t)$,
i.e., the displacements, velocities, and accelerations
of all floors of the left and right buildings.
The parameters used in this calculation
(and also in the rest of the paper) are as follows.
The dimensions are as shown in Fig.~\ref{fig_bldg}.
The mass and stiffness are 64,719 Kg and $3.7774\times 10^8$\,N/m
per storey. The VE damper properties are
$K_d \,$=$\, 10^6$\,N/m,
$C_d \,$=$\, 10^8$\,N-m/sec.

\begin{figure*}[!ht]
\centering
\scalebox{0.9}{\includegraphics{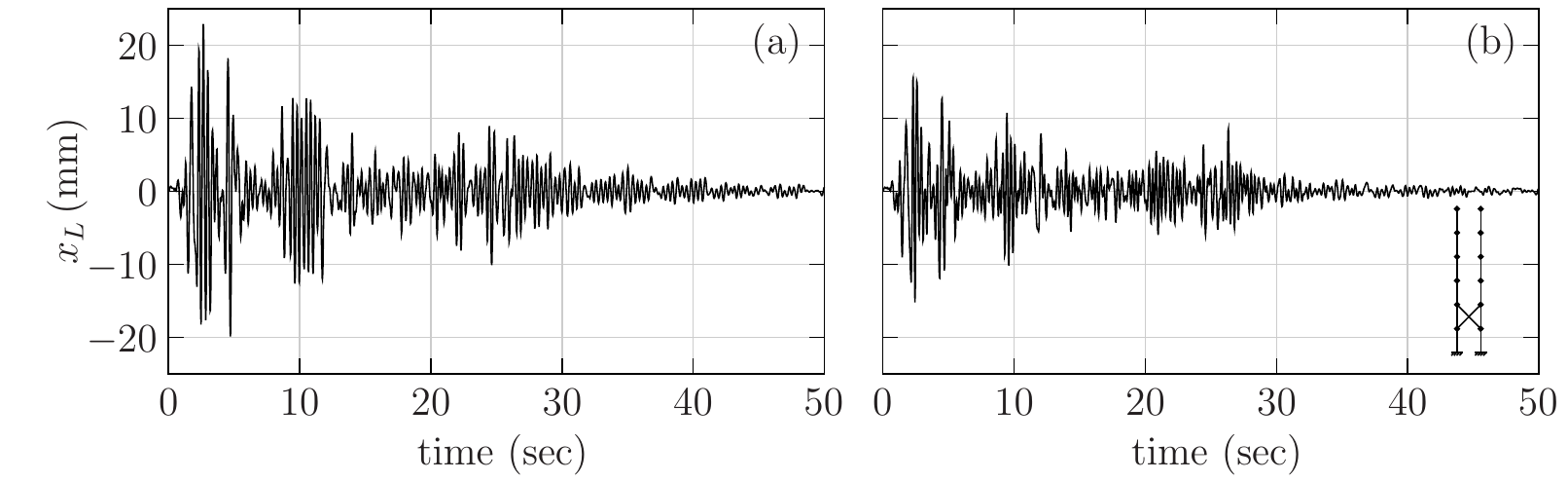}}
\vspace*{-0.2cm}
\caption{Displacement of top left floor versus time for six-storeyed
DSABs as obtained by solving Eq.~\ref{eq_motion}. (a)\,without
dampers, (b)\,with dampers connected as shown in the inset.}
\label{fig_sample_disp}
\end{figure*}

The top left floor displacement $x_L(t)$ obtained by solving
Eq.~\ref{eq_motion} for six-storeyed DSABs is shown in
Fig.~\ref{fig_sample_disp}\,(a)
when no dampers are used.
Fig.~\ref{fig_sample_disp}\,(b)
shows $x_L(t)$ for the same DSABs but with two dampers
connected as shown in the figure. The reduction obtained in
$x_L(t)$ with the use of dampers is clearly seen.

There are several other possible locations for the dampers. The question
which naturally arises is whether there is an optimum placement which would
lead to smaller displacements (or improved values for any other objective
functions). Specifically, in this paper, we consider optimisation of the
damper configuration for a given number of floors ($N_f$) and a given
number of dampers ($N_d$) with the building and damper parameters, as
described above, held constant.

This paper is organised as follows. In
Sec.~\ref{sec_sols}, we describe the solution space for specific examples. In
Sec.~\ref{sec_problem}, we discuss the formulation of the multi-objective
optimisation problem considered in this work and present results obtained
for DSABs with $N_f \,$=$\, 10$ (ten storeys) and different values of
$N_d$. In
Sec.~\ref{sec_compare}, we compare the performance of two commonly used
multi-objective evolutionary algorithms (MOEAs), viz,
NSGA-II\,\cite{deb2002} and MOPSO\,\cite{coello2004} in the context of
the damper placement optimisation problem. Finally, in
Sec.~\ref{sec_conclusions}, we present conclusions of this work along
with some future research directions.

\section{Solution space examples}
\label{sec_sols}
The solution space for a multi-objective optimisation problem depends on
the choice of the objective functions. For the damper placement optimisation
problem, various objective functions have been used in the literature\,\cite{hadi2015}.
Here, we describe the solution space for three sets of objective functions
for six-storeyed DSABs (i.e., $N_f \,$=$\, 6$) with two or three dampers
(i.e., $N_d \,$=$\, 3$ or $N_d \,$=$\, 4$).

\subsection{Maximum floor displacements}
\label{sec_sols_1}
\begin{figure*}[!ht]
\centering
\scalebox{0.9}{\includegraphics{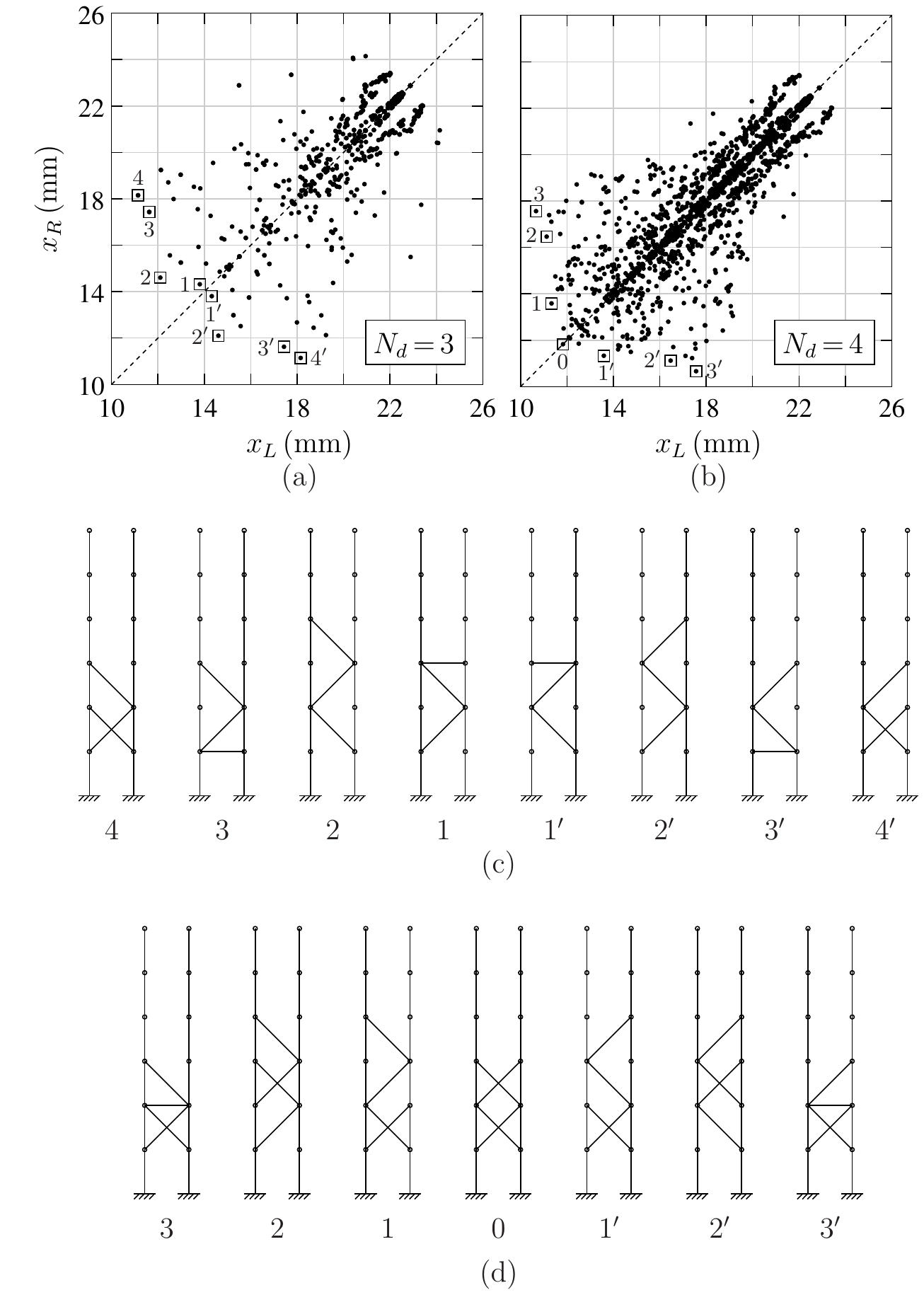}}
\caption{Maximum top left floor and top right floor displacements
for six-storeyed DSABs.
(a)\,solutions space for $N_d \,$=$\, 3$,
(b)\,solutions space for $N_d \,$=$\, 4$,
(c)\,damper placement for Pareto-optimal solutions with $N_d \,$=$\, 3$,
(d)\,damper placement for Pareto-optimal solutions with $N_d \,$=$\, 4$.}
\label{fig_solspace1}
\end{figure*}

In this case, we consider two objective functions, viz., the maximum
displacements (over time) of the top left floor and the top
right floor, denoted by $x_L$ and $x_R$, respectively.
Figs.~\ref{fig_solspace1}\,(a) and
\ref{fig_solspace1}\,(b)
show the solution space (i.e., solutions obtained for all
possible damper configurations) for
$N_d \,$=$\, 3$ and $N_d \,$=$\, 4$.
The Pareto-optimal solutions are marked with squares.
Since the two buildings are dynamically similar, the
$x_L$ and $x_R$
values are symmetrically located, as seen in the figure.
Figs.~\ref{fig_solspace1}\,(c) and
\ref{fig_solspace1}\,(d)
show the damper configurations corresponding to the Pareto-optimal
solutions.

The total number of possible damper positions for
$N_f \,$=$\, 6$ is
$N_f + 2\,(N_f-1) \,$=$\, 16$, as shown in
Fig.~\ref{fig_bldg}\,(c), and
the total number of damper configurations with
$N_d \,$=$\, 3$
is\,\cite{bigdeli2012} $^{16} C _3 \,$=$\, 560$.
The solution space in
Fig.~\ref{fig_solspace1}\,(a)
has been obtained by solving Eq.~\ref{eq_motion} for each of these 560
configurations and recording $x_L$ and $x_R$ in each case.
It may be noted that some of the
$(x_L,x_R)$ values would overlap, and therefore the number of distinct
solutions in
Fig.~\ref{fig_solspace1}\,(a)
would be less than 
$^{16} C _3$. Once the entire solution space is found, it is a straightforward
matter to compare the solutions with each other to obtain the Pareto front.

When the number of possible configurations becomes much larger,
it would take too long to evaluate all of them.
Furthermore, as $N_f$ increases, the matrix sizes in Eq.~\ref{eq_motion}
also increase, which means that the solution time per configuration is also larger.
For this reason, the above ``enumeration" or ``exhaustive search"\,\cite{bigdeli2012}
procedure becomes impractical for many cases of practical interest. In this scenario,
MOEAs~-- which would generally require a much smaller number of function evaluations~--
provide an attractive alternative.

\subsection{Maximum acceleration and interstorey drift}
\label{sec_sols_2}
\begin{figure*}[!ht]
\centering
\scalebox{0.9}{\includegraphics{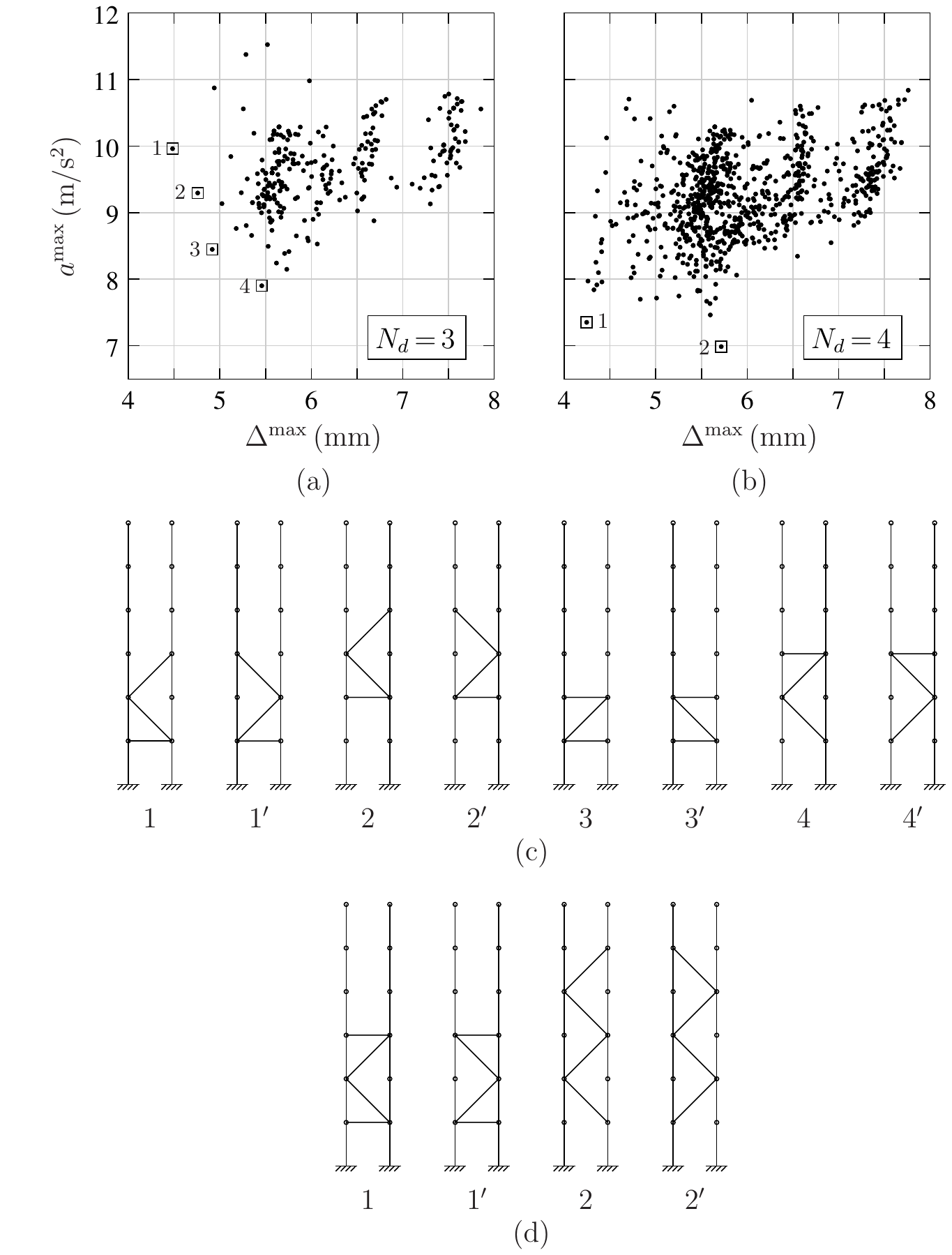}}
\caption{Maximum acceleration and maximum interstorey drift (over all
floors and time) for six-storeyed DSABs.
(a)\,solutions space for $N_d \,$=$\, 3$,
(b)\,solutions space for $N_d \,$=$\, 4$,
(c)\,damper placement for Pareto-optimal solutions with $N_d \,$=$\, 3$,
(d)\,damper placement for Pareto-optimal solutions with $N_d \,$=$\, 4$.}
\label{fig_solspace2}
\end{figure*}

We now consider another set of two objective functions, viz.,
the maximum acceleration
$a^{\mathrm{max}}$
and the maximum interstorey drift\,\cite{aydin2013}
$\Delta^{\mathrm{max}}$
(over all floors and time).
Figs.~\ref{fig_solspace2}\,(a) and
\ref{fig_solspace2}\,(b)
show the solutions space for
$N_d \,$=$\, 3$ and $N_d \,$=$\, 4$, respectively.
The damper configurations corresponding to the Pareto-optimal solutions are shown
Figs.~\ref{fig_solspace2}\,(c) and
\ref{fig_solspace2}\,(d).

\subsection{Maximum top-floor displacement and base shear}
\label{sec_sols_3}
\begin{figure*}[!ht]
\centering
\scalebox{0.9}{\includegraphics{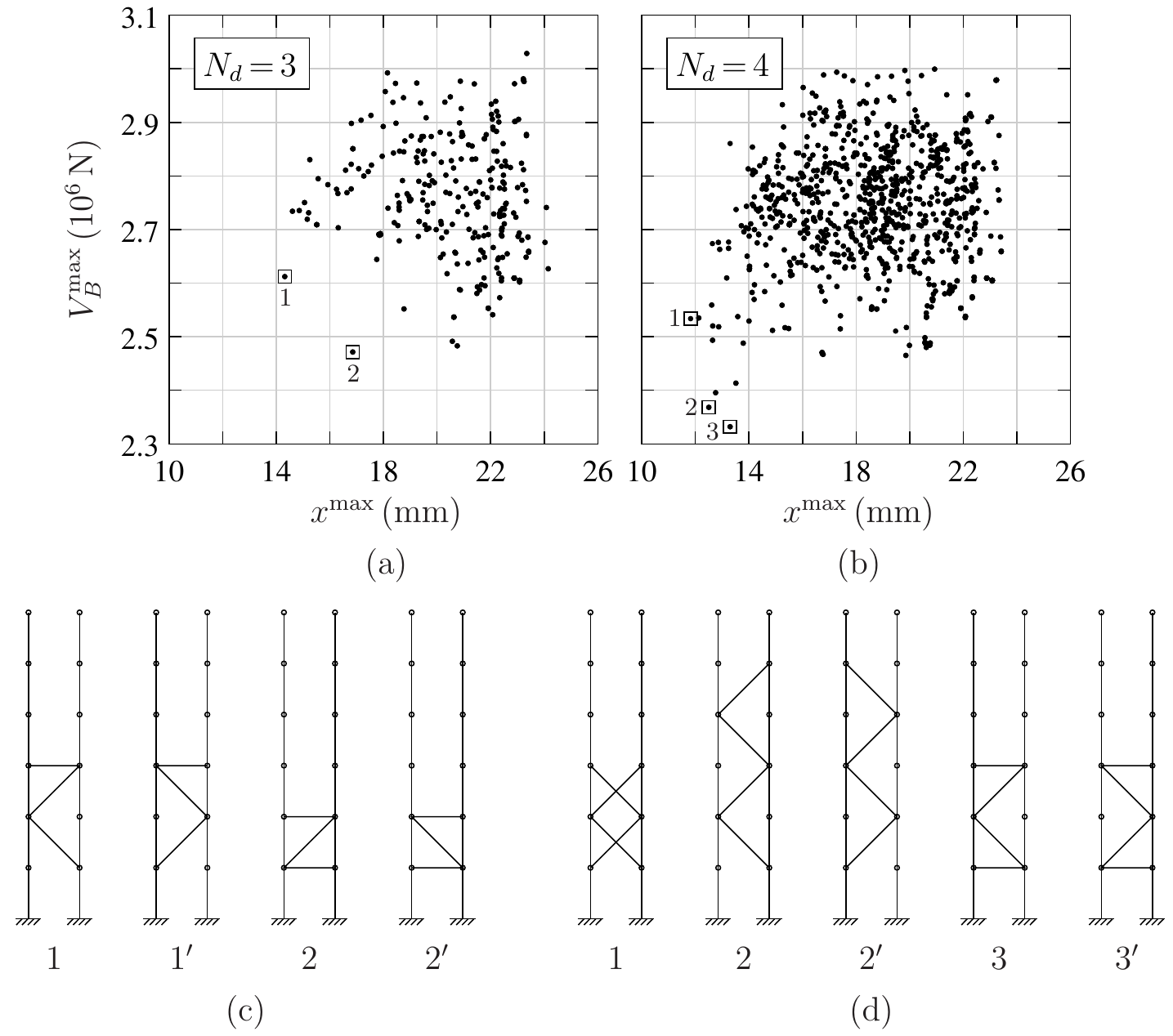}}
\caption{Maximum top-floor displacement (over all
floors and time) and maximum base shear (over time) for six-storeyed DSABs.
(a)\,solutions space for $N_d \,$=$\, 3$,
(b)\,solutions space for $N_d \,$=$\, 4$,
(c)\,damper placement for Pareto-optimal solutions with $N_d \,$=$\, 3$,
(d)\,damper placement for Pareto-optimal solutions with $N_d \,$=$\, 4$.}
\label{fig_solspace3}
\end{figure*}

The maximum base shear is yet another quantity of interest\,\cite{bhaskararao2006}.
The solution space, with the maximum top-floor displacement
(over all floors and time) and the maximum base shear (over time) as the two
objectives, is shown in
Figs.~\ref{fig_solspace3}\,(a) and
\ref{fig_solspace3}\,(b)
for the same $N_f$ and $N_d$ values as before. The damper
configurations corresponding to the Pareto-optimal solutions are shown in
Figs.~\ref{fig_solspace3}\,(c) and
\ref{fig_solspace3}\,(d).

From
Figs.~\ref{fig_solspace1}-\ref{fig_solspace3}, we observe that
the three Pareto fronts have some damper
configurations in common. For example, configurations $1$ and $1'$ of
Fig.~\ref{fig_solspace1}\,(c)
also appear in
Figs.~\ref{fig_solspace2}\,(c) and \ref{fig_solspace3}\,(c). However, there
are other configurations which do not appear in all three Pareto fronts.

The most striking feature of the solution spaces shown in
Figs.~\ref{fig_solspace1}-\ref{fig_solspace3}
is the very small number of solutions in the Pareto front.
Typically, the performance of an MOEA is evaluated with measures
such as generational distance, error ratio, and spread\,\cite{kdeb1}.
For the damper placement optimisation problem considered here, these
measures are clearly not applicable. Instead, the most relevant performance
measure for this problem is the number of times the MOEA is able to obtain the
Pareto-optimal solutions in a given number of independent runs. We will quantify
this performance measure in terms of the ``success rate" (SR) for the 
$k^{\mathrm{th}}$ Pareto-optimal solution as
\begin{equation}
SR\,(k) = \displaystyle\frac{N_k}{N_r},
\label{eq_sr}
\end{equation}
where $N_r$ is the number of independent runs (trials) of the MOEA,
and $N_k$ is the number of runs in which the
$k^{\mathrm{th}}$ Pareto-optimal solution was obtained. We will use the
above definition to compare MOEAs with respect to the damper placement
problem in Sec.~\ref{sec_compare}.

\section{Problem formulation}
\label{sec_problem}
The multi-objective optimisation problem considered in this paper can be
stated as follows. Given two DSABs with $N_f$ floors and $N_d$ identical
dampers, find the Pareto-optimal set of solutions (damper configurations)
with the objectives of minimising $f_1$ and $f_2$. We consider two sets
of $(f_1,\,f_2)$\,:
\begin{list}{(\roman{cntr2})}{\usecounter{cntr2}}
 \item
  $f_1$ is the maximum interstorey drift
  ($\Delta ^{\mathrm{max}}$),
  and $f_2$ is the maximum acceleration
  ($a^{\mathrm{max}}$), where
  $\Delta ^{\mathrm{max}}$ and
  $a^{\mathrm{max}}$
  are maximum values over time and over all floors.
  As shown in \cite{lavan2009}, these two objectives are conflicting in nature, thus
  making it a multi-objective optimisation problem.
 \item
  $f_1$ is the maximum top-floor displacement
  ($x ^{\mathrm{max}}$),
  and $f_2$ is the maximum base shear\,\cite{bhaskararao2006}
  ($V_B^{\mathrm{max}}$), where
  $x ^{\mathrm{max}}$
  and $V_B^{\mathrm{max}}$ are maximum values over time.
\end{list}
Note that several other objective functions have been used in the literature\,\cite{hadi2015}.
Here, we have selected two representative sets of objectives.
Our main focus is on the optimisation issues involved, and the conclusions drawn from
this work are expected to be broadly applicable
for other choices of objective functions as well.

Applying an MOEA to the above problem involves repeated evaluation
of the objective functions
for specific damper configurations. If the NSGA-II algorithm\,\cite{deb2002}
is used, each chromosome in the population would correspond to a specific damper
configuration.
\begin{figure}[!ht]
\centering
\scalebox{0.6}{\includegraphics{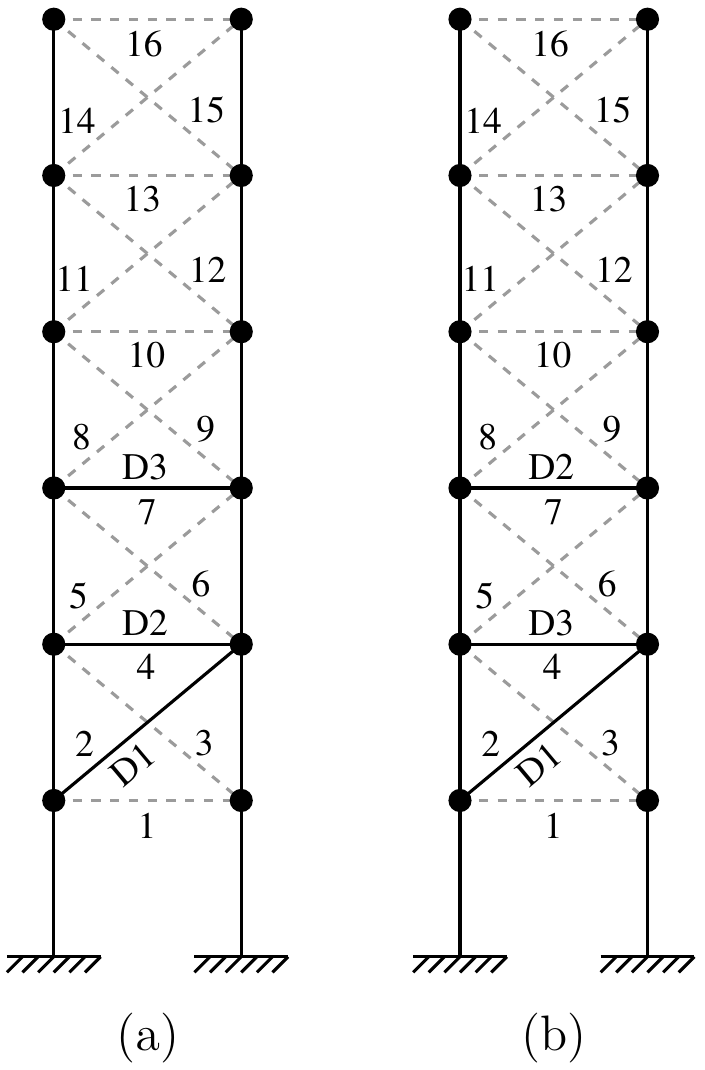}}
\caption{Damper configuration examples for $N_f \,$=$\, 6$, $N_d \,$=$\, 3$.}
\label{fig_config}
\end{figure}
As an example, consider the damper configuration shown in
Fig.~\ref{fig_config}\,(a).
Treating each damper position $d_k$ as a decision variable, the
chromosome representing this configuration would be characterised by
$d_1 \,$=$\, 2$,
$d_2 \,$=$\, 4$,
$d_3 \,$=$\, 7$.
Similarly, if the multi-objective particle swarm optimisation (MOPSO)
algorithm\,\cite{coello2004} is used, the particle representing the
configuration in
Fig.~\ref{fig_config}\,(a)
would have three parameters, viz.,
$d_1 \,$=$\, 2$,
$d_2 \,$=$\, 4$,
$d_3 \,$=$\, 7$.

Since we have assumed the three dampers to have identical properties,
the configurations in
Figs.~\ref{fig_config}\,(a) and
\ref{fig_config}\,(b)
would give the same objective function values although their decision
variable values differ
($d_1 \,$=$\, 2$,
$d_2 \,$=$\, 4$,
$d_3 \,$=$\, 7$ in Fig.~\ref{fig_config}\,(a), and
$d_1 \,$=$\, 2$,
$d_2 \,$=$\, 7$,
$d_3 \,$=$\, 4$ in Fig.~\ref{fig_config}\,(b)).
Clearly, it is wasteful to evaluate both of these configurations.
In order to avoid such repetitive computations, we evaluate a chromosome
(or particle) only if the damper positions satisfy
$d_3 > d_2 > d_1$. If this condition is not satisfied, we assign suitably
large values (``penalties") to the objective functions $f_1$ and $f_2$, thus making that
solution unfit.

Another feature which can be used to limit the search space is the
following. For
$N_f \,$=$\, 6$ and
$N_d \,$=$\, 3$,
there are a total of 16 possible damper positions, as shown in
Fig.~\ref{fig_bldg}\,(c). The first position
cannot be occupied by D2 or D3 since D1 needs to occupy a position lower than D2.
Similarly, the $16^{\mathrm{th}}$ position in the figure cannot be occupied
by D1 or D2 since D3 needs to occupy a position higher than D2. With this logic,
we can limit the search space by restricting the decision variables as\\
\begin{eqnarray}
\nonumber
d_1 \!\!& \in & \!\! \{1,2,\cdots ,13,14\},\\
d_2 \!\!& \in & \!\! \{2,3,\cdots ,14,15\},\\
d_3 \!\!& \in & \!\! \{3,4,\cdots ,15,16\}.
\label{eq_limit}
\end{eqnarray}
Although the decision variables ($d_k$) in our problem take on only
integer values, we consider them as real variables and convert them
to integers before evaluating the objective functions.
We will describe further details of the algorithms used in this work in
Sec.~\ref{sec_compare}. 

\begin{figure*}[!ht]
\centering
\scalebox{0.9}{\includegraphics{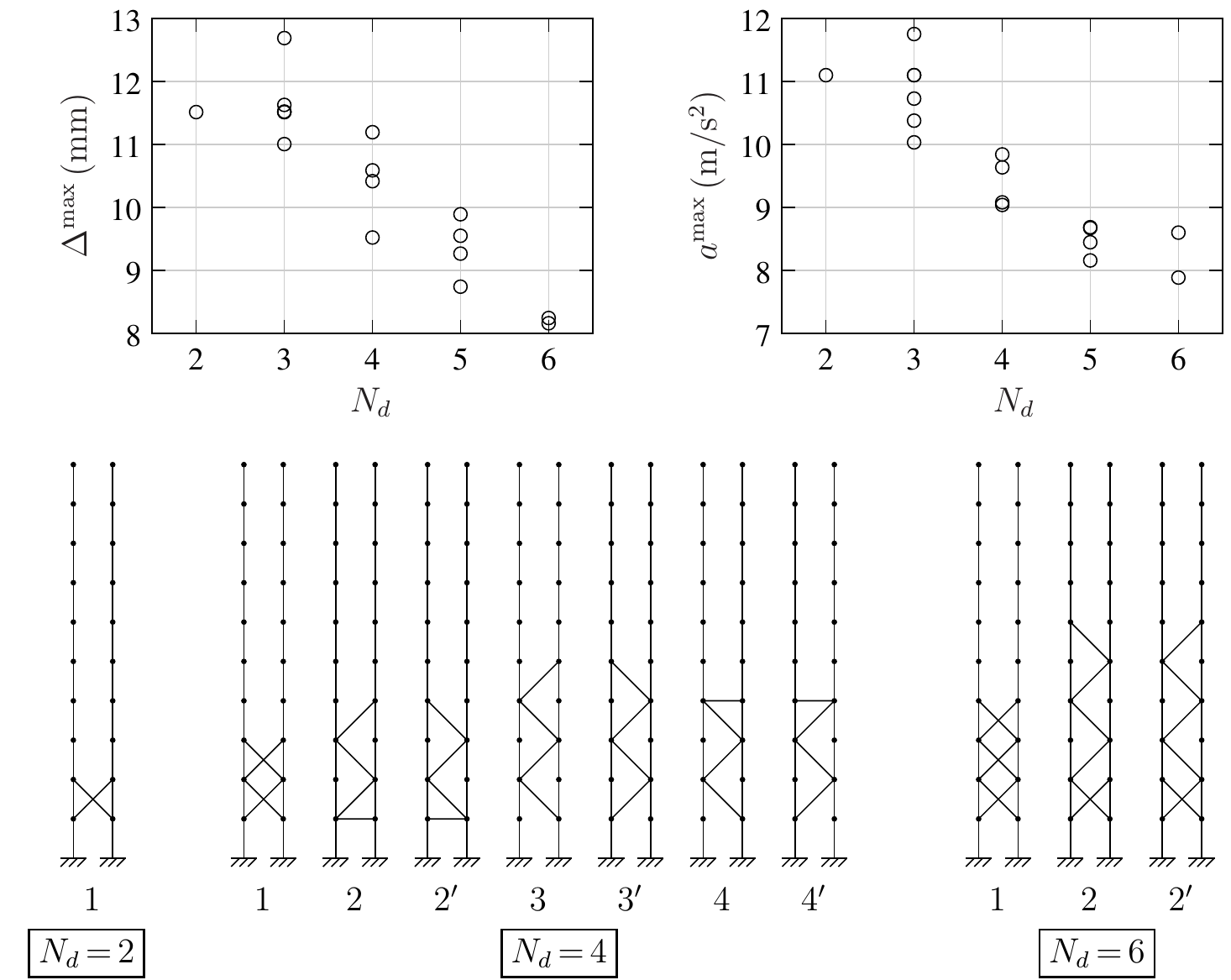}}
\caption{Objective function values for Pareto-optimal solutions
of the $(\Delta ^{\mathrm{max}},\,a^{\mathrm{max}})$ optimisation problem
for different values of $N_d$, and optimal damper configurations for
$N_d \,$=$\, 2$, 4, 6.}
\label{fig_nf10y}
\end{figure*}

We now look at the Pareto-optimal solutions obtained for $N_f \,$=$\, 10$
with different values of $N_d$.
Fig.~\ref{fig_nf10y}
shows the objective function values for the first set of objectives, i.e.,
$(\Delta ^{\mathrm{max}},\,a^{\mathrm{max}})$,
for the Pareto-optimal solutions for
$N_d \,$=$\, 2$, 3, 4, 5, 6, and the damper configurations for
$N_d \,$=$\, 2$, 4, 6.
\begin{table}
   \centering
   \caption{Pareto-optimal solutions for the
    $(\Delta ^{\mathrm{max}},\,a^{\mathrm{max}})$ optimisation problem, with
    $N_f \,$=$\, 10$ and $N_d \,$=$\, 2$, 4, 6. Numbers in the index column
    correspond to the damper configurations shown in Fig.~\ref{fig_nf10y}.}
    \hspace*{0cm}
    \vspace*{0.2cm}

\begin{tabular}{|c|c|r|r|}
  \hline
    {\multirow{2}{*}{$N_d$}}
  & {\multirow{2}{*}{index}}
  & \multicolumn{1}{c|}{{\rule{0pt}{3.0ex}}$\Delta _{\mathrm{max}}$}
  & \multicolumn{1}{c|}{$a_{\mathrm{max}}$}
  \\
    {}
  & {}
  & \multicolumn{1}{c|}{{\rule[-1.4ex]{0pt}{0pt}}(mm)}
  & \multicolumn{1}{c|}{(m/s$^2$)}
\\ \hline
{\multirow{1}{*}{2}}
& \TTT 1 & 11.52 & 11.10
\\ \hline
  {\multirow{4}{*}{4}}
  & \TTT 1 & 9.52 & 9.84
  \\ \cline{2-4}
  {}
  & \TTT $2$,\,$2'$ & 10.42 & 9.64
  \\ \cline{2-4}
  {}
  & \TTT $3$,\,$3'$ & 10.59 & 9.08
  \\ \cline{2-4}
  {}
  & \TTT $4$,\,$4'$ & 11.20 & 9.04
\\ \hline
  {\multirow{2}{*}{6}}
  & \TTT 1 & 8.16 & 8.60
  \\ \cline{2-4}
  {}
  & \TTT 2, 3 & 8.25 & 7.89
 
\\ \hline
\end{tabular}
\label{tbl_nf10y}
\end{table}
The objective function values for
$N_d \,$=$\, 2$, 4, 6 are also listed in Table~\ref{tbl_nf10y}.
We observe that, as the number of dampers increases, there is an overall
improvement in the objective function values. From Table~\ref{tbl_nf10y}, it is
also clear that the two objectives,
$a^{\mathrm{max}}$ and
$\Delta ^{\mathrm{max}}$,
are conflicting. As a result, the best configuration for minimum
$a^{\mathrm{max}}$ is in general different than that for minimum
$\Delta ^{\mathrm{max}}$, the only exception being the $N_d \,$=$\, 2$ case.

From Fig.~\ref{fig_nf10y}, we observe that the optimum configurations take
various geometric forms, thus making it difficult to generalise.
The complexity would increase further as more objective functions are considered
or additional decision variables, such as damper properties, are allowed.
For this reason, the use if MOEAs is the only practical option for choosing a
damper configuration when the solution space is large.

\begin{figure}[!ht]
\centering
\scalebox{0.9}{\includegraphics{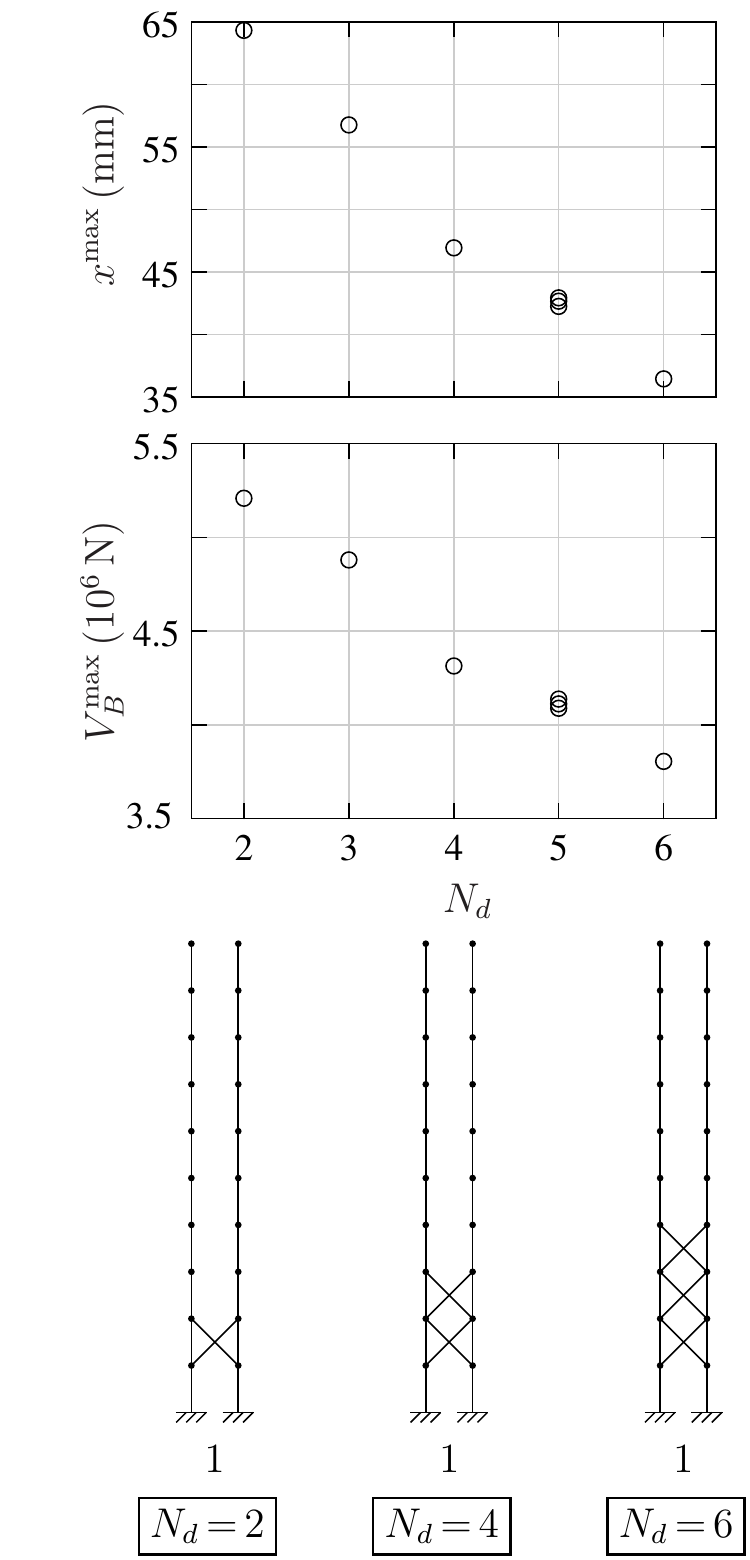}}
\caption{Objective function values for Pareto-optimal solutions
of the $(x ^{\mathrm{max}},\,V_B^{\mathrm{max}})$ optimisation problem
for different values of $N_d$, and optimal damper configurations for
$N_d \,$=$\, 2$, 4, 6.}
\label{fig_nf10a}
\end{figure}

The Pareto-optimal objective function values for the second set, i.e.,
$(x^{\mathrm{max}},\,V_B^{\mathrm{max}})$, are shown
in Fig.~\ref{fig_nf10a}. The optimal damper configurations
for $N_d \,$=$\, 2$, 4, 6
are also shown in the figure, and the corresponding objective function
values are listed in
Table~\ref{tbl_nf10a}.
For these three cases, the configuration which results in minimum
$x^{\mathrm{max}}$
also gives the minimum
$V_B^{\mathrm{max}}$.
These ``criss-cross" configurations also appear in the Pareto-optimal sets for the first
objective set
(Fig.~\ref{fig_nf10y}) and have been found to be beneficial in the study presented
in \cite{patel2010} as well.

\begin{table}
   \centering
   \caption{Pareto-optimal solutions for the
    $(x ^{\mathrm{max}},\,V_B^{\mathrm{max}})$ optimisation problem, with
    $N_f \,$=$\, 10$ and $N_d \,$=$\, 2$, 4, 6. Numbers in the index column
    correspond to the damper configurations shown in Fig.~\ref{fig_nf10a}.}
    \hspace*{0cm}
    \vspace*{0.2cm}

\begin{tabular}{|c|c|r|r|}
  \hline
    {\multirow{2}{*}{$N_d$}}
  & {\multirow{2}{*}{index}}
  & \multicolumn{1}{c|}{{\rule{0pt}{3.0ex}}$x ^{\mathrm{max}}$}
  & \multicolumn{1}{c|}{$V_B ^{\mathrm{max}}$}
  \\
    {}
  & {}
  & \multicolumn{1}{c|}{{\rule[-1.4ex]{0pt}{0pt}}(mm)}
  & \multicolumn{1}{c|}{($10^6$\,N)}
\\ \hline
{\multirow{1}{*}{2}}
& \TTT 1 & 64.31 & 5.21
\\ \hline
  {\multirow{1}{*}{4}}
  & \TTT 1 & 46.92 & 4.31
\\ \hline
  {\multirow{1}{*}{6}}
  & \TTT 1 & 36.43 & 3.81
 
\\ \hline
\end{tabular}
\label{tbl_nf10a}
\end{table}

\section{Comparison of MOEAs}
\label{sec_compare}
As the solutions space grows, the efficiency of the MOEA becomes more
important. Let us take a specific example, viz.,
$N_f \,$=$\, 10$,
$N_d \,$=$\, 6$.
In this case, the total number of possible damper positions is
$N_f + 2\,(N_f-1) \,$=$\, 28$, and 
the size of the solution space $S$ (defined as the total number of damper configurations)
is $^{28} C _6 \,$=$\, 376,740$.
The average CPU time required for evaluating one configuration was found to be about
6.6\,msec
on a desktop computer (Linux)
with 3.8\,GHz clock and 8\,GB RAM without any parallelization.
Exhaustive search, i.e., evaluation of all possible configurations, would take
$376,740 \times 6$\,msec or about 41 minutes.
With large values of $N_f$ or $N_d$, the exhaustive search option becomes too expensive,
and therefore an MOEA, which can obtain the Pareto front solutions by evaluating a small fraction
of $S$, is certainly desirable. With this in mind, we define the following two figures
of merit for an MOEA in the context of damper placement optimisation for DSABs.
\begin{list}{(\alph{cntr2})}{\usecounter{cntr2}}
 \item
  Computational effort
  $CE \,$=$\, \displaystyle\frac{N_{FE}^{\mathrm{total}}}{S}$, where
  $N_{FE}^{\mathrm{total}}$ is the total number of function evaluations carried out
  by the MOEA in $N_r$ independent runs, and $S$ is the size of the solution space.
  Note that exhaustive search, which gives all Pareto-optimal solutions, requires
  $CE \,$=$\, 1$.
  An MOEA which requires
  $CE > 1$
  for obtaining all Pareto-optimal solutions
  is therefore of no practical use.
 \item
  Success rate $SR\,(k)$, as defined by Eq.~\ref{eq_sr}. An ideal MOEA would find
  all Pareto-optimal solutions in each run, thus giving
  $SR\,(k) \,$=$\, 1$ for each value of $k$, i.e., for each Pareto-optimal solution.
  In practice, $SR\,(k)$ would be smaller.
\end{list}
With the above definitions, we can now compare MOEAs with each other.
For the same $CE$, an MOEA with a larger
$SR\,(k)$
is better. Similarly, for the same
$SR\,(k)$,
an MOEA with a smaller $CE$ is better. In the following, we will consider the
damper placement problem with
$N_f \,$=$\, 10$ and
$N_d \,$=$\, 6$, and compare the performance of three MOEAs, as described below.
\begin{list}{(\alph{cntr2})}{\usecounter{cntr2}}
 \item
  NSGA-II: This is the real-coded
  nondominated sorting genetic algorithm presented in \cite{deb2002}.
  The following algorithm
  parameters were chosen:
  crossover probability $p_c \,$=$\, 0.9$,
  distribution index for crossover $\eta _c \,$=$\, 15$,
  mutation probability $p_{\mathrm{mut}} \,$=$\, 1/L$ (where $L \,$=$\, N_d$ is the number
  of decision variables), and
  distribution index for mutation $\eta _m \,$=$\, 7$.
  Since 
  $p_{\mathrm{mut}} \,$=$\, 1/N_d$,
  on average, one of the damper positions of a given chromosome get mutated.
  The following polynomial probability distribution is used for finding the
  mutated parameter value\,\cite{deb2002}.
  \begin{equation}
  P\,(\delta) = 0.5\,\left(\eta _m + 1\right)\left(1-|\,\delta \,|\,\right)^{\eta _m}.
  \label{eq_poly}
  \end{equation}
  Eq.~\ref{eq_poly} ensures that the new (mutated) parameter value is
  close to the previous value.
 \item
  MOPSO-1: This is the same as the MOPSO algorithm described in \cite{coello2004}.
  The algorithm parameters used in this work are
  $W \,$=$\, 0.4$,
  $C_1 \,$=$\, 2$,
  $C_2 \,$=$\, 2$.
  As described in
  \cite{coello2004},
  the mutation operator is implemented as follows. For example, consider
  a constant probability of mutation
  $p_{\mathrm{mut}} \,$=$\, 0.05$. In this case, in each PSO iteration,
  5\,\% of the particles (on average) are randomly selected, and for each
  of them, one of the parameters, i.e., the position of one of the $N_d$
  dampers, is randomly mutated.
 \item
  MOPSO-2: This is identical to MOPSO-1 except for the implementation of the
  mutation operation. In MOPSO-2, for each particle selected for mutation,
  one of the parameters is mutated, as in MOPSO-1. However, instead of mutating
  it randomly, the polynomial probability distribution given by Eq.~\ref{eq_poly}
  is used, with $\eta _m \,$=$\, 7$.
\end{list}
Several choices exist for the algorithm parameters
(such as $p_c$, $\eta _c$, $\eta _m$ for NSGA-II
and $W$, $C_1$, $C_2$ for MOPSO). Here, we have selected parameter
values which have been found to be effective in the literature. Apart from
the algorithm parameters, the population size $N_p$ (number of chromosomes
in NSGA-II or number of particles in MOPSO), the number of iterations
$N_{\mathrm{iter}}$, and the number of independent runs $N_r$ also play a role
in the performance of an MOEA (e.g., see \cite{coello2004}). In this work, we have
varied
$N_p$,
$N_{\mathrm{iter}}$,
$N_r$,
and recorded the performance measures in each case. The results are shown in
Tables~\ref{tbl_nf10nd6y} and \ref{tbl_nf10nd6a} for the
$(\Delta ^{\mathrm{max}},\,a^{\mathrm{max}})$
and $(x ^{\mathrm{max}},\,V_B^{\mathrm{max}})$ optimisation problems, respectively.
Note that there are three
success rates in
Table~\ref{tbl_nf10nd6y},
corresponding to the three
Pareto-optimal solutions shown in Fig.~\ref{fig_nf10y} ($N_d \,$=$\, 6$ case).
For example, for the MOPSO-1 algorithm with
$N_p \,$=$\, 40$,
$N_{\mathrm{iter}} \,$=$\, 200$,
$N_r \,$=$\, 30$,
$SR\,(1) \,$=$\, 4/30$, indicating that solution
1 has been found 4 times in 30 independent runs.
\begin{table*}
   \centering
   \caption{Summary of performance of NSGA-II, MOPSO-1, and MOPSO-2 algorithms
    for the
    $(\Delta ^{\mathrm{max}},\,a^{\mathrm{max}})$ optimisation problem, with
    $N_f \,$=$\, 10$ and
    and $N_d \,$=$\, 6$.}
    \hspace*{0cm}
    \vspace*{0.2cm}

\begin{tabular}{|c|r|r|r|r|r|r|r|r|}
  \hline
    Algorithm
  & \multicolumn{1}{c|}{{\rule{0pt}{3.0ex}}$N_p$}
  & \multicolumn{1}{c|}{{\rule{0pt}{3.0ex}}$N_{\mathrm{iter}}$}
  & \multicolumn{1}{c|}{{\rule{0pt}{3.0ex}}$N_r$}
  & \multicolumn{1}{c|}{{\rule{0pt}{3.0ex}}$P_{\mathrm{mut}}$}
  & \multicolumn{1}{c|}{CE}
  & \multicolumn{1}{c|}{SR\,(1)}
  & \multicolumn{1}{c|}{SR\,(2)}
  & \multicolumn{1}{c|}{SR\,(3)}
\\ \hline
  {\multirow{4}{*}{NSGA-II}}
  & \TTT 40 & 200 & 30 & $1/N_d$ & ~0.637~ & 0/30~ & 2/30~ & 4/30~
  \\ \cline{2-9}
  {}
  & \TTT 40 & 400 & 30 & $1/N_d$ & ~1.274~ & 0/30~ & 2/30~ & 5/30~
  \\ \cline{2-9}
  {}
  & \TTT 100 & 60 & 10 & $1/N_d$ & ~0.159~ & 0/10~ & 1/10~ & 3/10~
  \\ \cline{2-9}
  {}
  & \TTT 100 & 120 & 10 & $1/N_d$ & ~0.319~ & 0/10~ & 1/10~ & 4/10~
\\ \hline
  {\multirow{4}{*}{MOPSO-1}}
  & \TTT 40 & 200 & 30 & 0.05 & ~0.637~ & 4/30~ & 4/30~ & 6/30~
  \\ \cline{2-9}
  {}
  & \TTT 40 & 400 & 30 & 0.05 & ~1.274~ & 4/30~ & 5/30~ & 9/30~
  \\ \cline{2-9}
  {}
  & \TTT 40 & 200 & 30 & 0.20 & ~0.637~ & 3/30~ & 9/30~ & 11/30~
  \\ \cline{2-9}
  {}
  & \TTT 40 & 400 & 30 & 0.20 & ~1.274~ & 4/30~ & 9/30~ & 13/30~
\\ \hline
  {\multirow{2}{*}{MOPSO-2}}
  & \TTT 100 & 60 & 10 & 0.05 & ~0.159~ & 2/10~ & 1/10~ & 3/10~
  \\ \cline{2-9}
  {}
  & \TTT 100 & 60 & 10 & 0.20 & ~0.159~ & 0/10~ & 3/10~ & 5/10~
 
\\ \hline
\end{tabular}
\label{tbl_nf10nd6y}
\end{table*}

\begin{table*}
   \centering
   \caption{Summary of performance of NSGA-II, MOPSO-1, and MOPSO-2 algorithms
    for the
    $(x ^{\mathrm{max}},\,V_B^{\mathrm{max}})$ optimisation problem, with
    $N_f \,$=$\, 10$ and
    and $N_d \,$=$\, 6$.}
    \hspace*{0cm}
    \vspace*{0.2cm}

\begin{tabular}{|c|r|r|r|r|r|r|}
  \hline
    Algorithm
  & \multicolumn{1}{c|}{{\rule{0pt}{3.0ex}}$N_p$}
  & \multicolumn{1}{c|}{{\rule{0pt}{3.0ex}}$N_{\mathrm{iter}}$}
  & \multicolumn{1}{c|}{{\rule{0pt}{3.0ex}}$N_r$}
  & \multicolumn{1}{c|}{{\rule{0pt}{3.0ex}}$P_{\mathrm{mut}}$}
  & \multicolumn{1}{c|}{CE}
  & \multicolumn{1}{c|}{SR\,(1)}
\\ \hline
  {\multirow{4}{*}{NSGA-II}}
  & \TTT 40 & 200 & 30 & $1/N_d$ & ~0.637~ & 0/30~
  \\ \cline{2-7}
  {}
  & \TTT 40 & 400 & 30 & $1/N_d$ & ~1.274~ & 3/30~
  \\ \cline{2-7}
  {}
  & \TTT 100 & 60 & 10 & $1/N_d$ & ~0.159~ & 0/10~
  \\ \cline{2-7}
  {}
  & \TTT 100 & 120 & 10 & $1/N_d$ & ~0.319~ & 0/10~
\\ \hline
  {\multirow{4}{*}{MOPSO-1}}
  & \TTT 40 & 200 & 30 & 0.05 & ~0.637~ & 3/30~
  \\ \cline{2-7}
  {}
  & \TTT 40 & 400 & 30 & 0.05 & ~1.274~ & 3/30~
  \\ \cline{2-7}
  {}
  & \TTT 40 & 200 & 30 & 0.20 & ~0.637~ & 4/30~
  \\ \cline{2-7}
  {}
  & \TTT 40 & 400 & 30 & 0.20 & ~1.274~ & 6/30~
\\ \hline
  {\multirow{2}{*}{MOPSO-2}}
  & \TTT 100 & 60 & 10 & 0.05 & ~0.159~ & 3/10~
  \\ \cline{2-7}
  {}
  & \TTT 100 & 60 & 10 & 0.20 & ~0.159~ & 0/10~
 
\\ \hline
\end{tabular}
\label{tbl_nf10nd6a}
\end{table*}
We can make the following observations from
Tables~\ref{tbl_nf10nd6y} and
\ref{tbl_nf10nd6a}.
\begin{list}{(\alph{cntr2})}{\usecounter{cntr2}}
 \item
  The NSGA-II algorithm consistently misses out one of the solutions in
  Table~\ref{tbl_nf10nd6y} even with $CE > 1$.
 \item
  From the first two rows of the NSGA-II section and the first two rows
  of the MOPSO-1 section of
  Table~\ref{tbl_nf10nd6y},
  we find that the success rate of NSGA-II has not changed significantly
  by doubling 
  $N_{\mathrm{iter}}$ whereas that of MOPSO-1 has improved.
 \item
  Compared to NSGA-II, the MOPSO algorithm is generally more effective
  in obtaining all Pareto-optimal solutions. This observation is similar
  to that in \cite{barraza2017} for a different optimisation problem.
 \item
  MOPSO-2, which involves the polynomial mutation operator, performs better
  than MOPSO-1. However, when the mutation probability $p_{\mathrm{mut}}$
  is increased from 0.05 to 0.2, it fails to capture one of the three solutions in
  Table~\ref{tbl_nf10nd6y}
  and the only solution in
  Table~\ref{tbl_nf10nd6a}.
\end{list}
The above discussion brings out the need for trying out different MOEAs,
possibly with some suitable changes incorporated in the algorithms,
for solving a given practical multi-objective optimisation problem.
In the literature, generally algorithms are evaluated for certain test
cases, and results are compared with other algorithms. Results of this
work suggest that different problems may require different algorithmic
options to be explored for improved efficiency.

\section{Conclusions}
\label{sec_conclusions}
Multi-objective optimisation of viscoelastic damper placement in DSABs
is presented. For six-storeyed DSABs with three and four dampers, the
complete solution spaces have been obtained for three sets of objectives.
In all three cases, it is shown that the number of solutions in the Pareto
front is very small. It is argued that, for the DSAB damper placement
problem, the figures of merit commonly used for evaluating MOEAs are not
meaningful. Two other figures of merit, viz., computational effort and
success rate, are therefore used in this work. For ten-storeyed DSABs,
optimum damper configurations have been obtained for different values
of $N_d$, the number of dampers. From the results obtained, it is concluded
that the use of MOEAs is the only effective way of obtaining optimum damper
placements when the solution space is large.

The performance of two commonly used MOEAs, viz., NSGA-II and MOPSO,
is compared in the context of the DSAB damper placement problem.
The MOPSO algorithm with a polynomial mutation operator is found to
be most effective.

The objective of this work was to mainly consider the optimisation aspects
of the DSAB damper placement problem. A number of related research directions
need to be investigated in future, as listed below.
\begin{list}{(\alph{cntr2})}{\usecounter{cntr2}}
 \item
  In this work, we have assumed all $N_d$ dampers to have identical parameters
  (stiffness and damping coefficients). These parameters could also be added to
  the decision variables to give greater flexibility to the decision maker.
 \item
  We have considered sets of two objective functions. For some problems, it
  may be desirable to consider three or four objective functions.
 \item
  A systematic study of the effect of $N_d$ on the best achievable objective
  values needs to be undertaken. As pointed out
  earlier\,\cite{bhaskararao2006},\cite{bigdeli2012},
  a smaller number of dampers with optimal parameters and positions may be
  adequate to meet the desired specifications. Optimisation can be used to
  confirm this finding.
 \item
  The present study could be extended to DSABs with larger $N_d$ and $N_f$.
\end{list}

\section*{Acknowledgement}

This work was partially supported by the Science and Engineering Board (SERB),
Department of science and technology (DST), Govt. of India. Financial support
from DST in the form of Fund for Improvement of S\&T Infrastructure (FIST)
is also gratefully acknowledged. M.B.P. would like to thank Prof. Kumar Appaiah,
IIT Bombay, for discussions related to programming aspects.

\bibliographystyle{IEEEtran}
\bibliography{damper1}

\end{document}